\documentclass[prc,aps,twocolumn]{revtex4}
\usepackage{graphicx} 
\usepackage{amsmath}
\usepackage{slashed}
\usepackage{ulem}
\usepackage{hyperref}
\usepackage{color}
\usepackage{url}

\begin{document}
 
 \title{QCD Evolution of Superfast Quarks}

\author{Adam~J.~Freese$^1$, Wim Cosyn$^2$ and Misak~M.~Sargsian$^3$}

\affiliation{
$^1$Physics Division, Argonne National Lab, Argonne, IL  60439 USA\\
$^2$Department of Physics and Astronomy, Ghent University, B9000 Ghent, Belgium\\
$^3$Department of Physics, Florida International University, Miami, FL 33199 USA}

\date{\today}

\begin{abstract} 
Recent high-precision  measurements of nuclear deep inelastic scattering at high $x$ and 
moderate $6 < Q^2 < 9$~GeV$^2$ give a rare  opportunity to reach the quark distributions in 
the {\it superfast} region,  in which the momentum fraction of the nucleon carried by its constituent 
quark is larger than the total fraction of the nucleon at rest, $x>1$. 
We derive the leading-order  QCD evolution equation for such  quarks  with the goal 
of relating the moderate-$Q^2$ data to the two earlier   measurements of superfast quark distributions 
at large $60 < Q^2 < 200$~GeV$^2$. Since  the high-$Q^2$  measurements gave strongly  
contradictory   estimates of the nuclear effects  that generate  superfast quarks, relating  them to the  
high-precision, moderate-$Q^2$ data  through  QCD evolution allows us to clarify this longstanding issue.
Our calculations indicate that  the moderate-$Q^2$ data at $x\lesssim 1.05$ are in better agreement with 
the  high-$Q^2$ data measured in (anti)neutrino-nuclear reactions   
which require substantial high-momentum nuclear effects in the generation of superfast quarks.
Our prediction for the high-$Q^2$ and $x>1.1$ region is somewhat in the 
middle of the  neutrino-nuclear and muon-nuclear scattering data.

\end{abstract}
\maketitle


\section{Introduction} 
With the operation of the Large Hadron Collider~(LHC), the high-energy upgrade of Jefferson Lab (JLab), and  the anticipation of the future electron-ion collider~(EIC),  
the  issue of understanding the partonic structure of nuclei  is  currently a very important topic. Several collaborations are working on the development of 
comprehensive parameterizations for nuclear partonic distributions (nPDFs) covering the widest  possible range of invariant momentum transfer $Q^2$ and
Bjorken variable $x$ (see. e.g.  Refs.\cite{Eskola:2016oht,Kovarik:2015cma,Owens:2012bv}). 

From the viewpoint of nuclear physics,   partons in nuclei present a very interesting dynamical construction as they are  constrained to be
in nucleons, which represent the apparent degrees of freedom in nuclei. Due to the large difference between the excitation energy scales 
of the nucleon~(100s of MeV) and  the nucleus  (10s of MeV),  it was initially believed that the nuclear medium  should play a non-essential role in 
the partonic dynamics  of  bound nucleons.  
Studies during the last several decades, however, discovered a host of effects which are genuinely related to  nuclear dynamics 
interfering with the QCD dynamics of partonic distributions in bound nucleons.  The most prominent of these effects is the 
suppression of nPDFs in the $0.4 < x< 0.7$ region (EMC effect)~\cite{Aubert:1983xm,Frankfurt:1988nt,Geesaman:1995yd}, nuclear anti-shadowing 
at $0.1 < x< 0.3$~\cite{Arneodo:1988aa,Frankfurt:1990xz,Brodsky:1989qz}, and finally the shadowing 
effects observed at $x<0.1$~\cite{Arneodo:1988aa,Frankfurt:2011cs,Frankfurt:1988nt}.

While there have been significant experimental and theoretical efforts in understanding the above mentioned effects, one effect which is 
less explored is the dynamics of superfast quarks.
Superfast quarks are quarks in nuclei possessing momentum fractions $x = \frac{A Q^2}{2 M_A q_0} > 1$ and
 represent one of the most elusive degrees of freedom in nuclei.  
Here $M_A$ is the mass of the nucleus A,  and $-Q^2$ and $q_0$ are the square of invariant momentum  transfer  and  the energy transferred 
to the nucleus in its rest frame. Since no such quark can be produced  by QCD dynamics  confined to a single nucleon  
without inter-nucleon interactions,  probing superfast quarks  requires direct interplay between QCD and 
nuclear dynamics.   One of the earliest theoretical studies of superfast quarks~\cite{Frankfurt:1988nt} showed that 
the nuclear dynamics responsible for the generation of such quarks is significantly short-range,   thus opening a new window 
into the high-density realm of  nuclear forces.  Such dynamics include multi-nucleon short-range 
correlations~\cite{Frankfurt:1988nt,Sargsian:2002wc,Freese:2014zda,Fomin:2017ydn},
explicit quark degrees of freedom such as 6-quark clusters~\cite{Pirner:1980eu,Carlson:1994ga}, or single-quark  momentum exchanges between 
strongly correlated nucleons~\cite{Sargsian:2007gd}. 

One way of probing superfast quarks experimentally is the extraction of the nuclear  structure function $F_{2A}(x,Q^2)$ 
in deep inelastic scattering~(DIS) from nuclei at $x>1$~\cite{Frankfurt:1988nt,Sargsian:2002wc,Pirner:1980eu}. Such studies are part of the physics program of the 12~GeV  
energy upgraded Jefferson Lab~\cite{AR_prop}.   Superfast quarks can also be probed in more unconventional processes such 
as semi-inclusive nuclear DIS processes  with tagged spectator nucleons~\cite{Melnitchouk:1996vp,Cosyn:2010ux,Cosyn:2017ekf},  
DIS production in the forward direction with $x_{F}>1$ or, large transverse momentum  dijet production  in  $p + A \rightarrow \mathrm{dijet} + X$ reactions  at LHC kinematics~\cite{Freese:2014zda}. All such processes will probe QCD dynamics in extreme nuclear conditions with the potential 
of opening up uncharted territory for nuclear QCD.

So far only three experiments have attempted to probe nuclear quark distributions at $x>1$. 
The first was carried out by the BCDMS collaboration at CERN~\cite{Benvenuti:1994bb}, which  measured the inclusive deep-inelastic scattering cross section on $^{12}$C at 
$52 \le Q^2 \le 200$~GeV$^2$. The second experiment was performed by the CCFR collaboration at Fermi Lab~\cite{Vakili:1999qt}, measuring 
neutrino and antineutrino charged current interactions from a $^{56}$Fe target at $\langle Q^2 \rangle= 125$~GeV$^2$.  Finally, the third experiment was 
performed more recently at Jefferson Lab~\cite{Fomin:2010ei}, where inclusive $A(e,e^\prime)X$ scattering cross section was measured at moderate 
values of   $6 \le Q^2 \le 9$~GeV$^2$. 

With the data of these experiments available, the main  motivation of our work is to investigate how these three results are related to each other
through the QCD evolution equation of nuclear partonic distribution functions. To carry out this study, we derive the QCD evolution equation for 
the nuclear structure function $F_{2A}$ and calculate the evolution of the Jefferson Lab data up to the $Q^2$ range of the BCDMS and CCFR experiments.

The outline of the paper is as follows: in Sec.~\ref{exp} we first give a brief  description of the available experiments     and quantify the existing discrepancy between the BCDMS and CCFR data.  Since the JLab data was taken at moderate values of $Q^2$, an important issue in the analysis in the high-$x$ region is the 
accounting of finite target mass~(TM) and higher twist~(HT) effects.  Therefore, the TM and HT corrections  procedure adopted by the JLab experiment is also described in Sec.~\ref{exp}.  
In Sec.~\ref{eveq}, we present the derivation of the QCD evolution equation for nuclear targets and obtain the self-consistent 
integro-differential equation for the nuclear structure function of $F_{2A}$. Then in Sec.~\ref{eveqnum}, the numerical solution of the evolution equation is obtained 
for the structure function parameterization obtained in Ref.~\cite{Fomin:2010ei} from the JLab data.  In Sec.~\ref{fit2}, we return to the issue of TM and HT corrections presenting  a different 
approach in accounting for these effects and presenting a new fit for the JLab $F_{2A}$ structure function. Our new fit indicates surprisingly small HT 
effects which we attribute to quark-hadron duality effects  amplified by the Fermi motion of bound nucleons in the nucleus.  
Our new fit does not alter the conclusion we obtained in Sec.~\ref{eveqnum} using the parametrization from Ref.~\cite{Fomin:2010ei}. 
However, it provides an improved description of 
the experimental data for $0.55 < x < 1.25$ over a wide $Q^2$ range.   For practical purposes in Sec.\ref{fit_of_fits} we  present a simple parameterization of 
the $F_{2A}$ parameters that allows estimation of the structure function over a wide range of $Q^2$ relevant to LHC and EIC kinematics.
In Sec.~\ref{NLO}, we check the accuracy of our calculations against next-to-leading 
order corrections, and finally Sec~.\ref{SumCon} states the summary and  conclusion of our work.

\medskip


\section{Experimental Evidence for Superfast Quarks}
\label{exp}
The first attempt to probe superfast quarks was made by the BCDMS collaboration~\cite{Benvenuti:1994bb}  in measuring the 
nuclear structure function $F_{2A}$ in deep-inelastic scattering of 200~GeV muons from a ${}^{12}$C  target.  The experiment
covered the region of $52 \le Q^2 \le 200$~GeV$^2$ and $x\le 1.3$, for the first time extracting the 
$F_{2A}$ structure function for $\langle Q^2 \rangle$ values of $61$, $85$ and $150$~GeV$^2$ at $x=0.85$, $0.95$, $1.05$, $1.15$, and
$1.30$. For these regions the per-nucleon $F_{2A}$   was fit to the form
\begin{equation}
F_{2A}(x,Q^2) = F_{2A}(x_0=0.75,Q^2)e^{-s(x-0.75)},
\label{slope}
\end{equation}
obtaining  $s=16.5\pm 0.6$ for the slope factor.
Such an exponent   required a larger  strength in  the high momentum 
distribution of nucleons in nuclei than the simple mean-field Fermi momentum distribution can provide. 
However  the amount of short-range correlations (that generate the high momentum strength) needed to agree with the data was very marginal.
 
The second experiment  was done by the CCFR collaboration~\cite{Vakili:1999qt} using neutrino and antineutrino beams 
and measured  the  per nucleon $F_{2A}$ structure function for ${}^{56}$Fe   in the charged current sector for 
$\langle Q^2 \rangle= 125$~GeV$^2$  and $0.6 \le x \le 1.2$. The experiment did not measure the absolute magnitudes of 
$F_{2A}$, but obtained the slope of the $x$ distribution in the form of Eq.~(\ref{slope}), with the exponent 
being evaluated as $s=8.3\pm 0.7 \pm 0.7$. This result was in clear contradiction with the  BCDMS result, requiring 
a much larger high-momentum component in the wave function of the ${}^{56}$Fe nucleus. 
The required high-momentum component was much larger than the one deduced from  quasi-elastic electroproduction 
in the $x>1$ region~\cite{Frankfurt:1993sp,Egiyan:2003vg,Egiyan:2005hs,Fomin:2011ng,Sargsian:2012sm,McGauley:2011qc}.

Recently, at JLab, the structure function $F_{2A}$ has been measured for a set of nuclei
(${}^2$H, ${}^3$He, ${}^4$He, ${}^9$Be, ${}^{12}$C, ${}^{63}$Cu, and ${}^{197}$Au) 
 over a wide range of $x$ (including $x>1$)  and  $Q^2$ ($2$-$9$~GeV$^2$)~\cite{Fomin:2010ei}. 
The  $F_{2A}$  extracted for the highest $Q^2$ ($6$-$9$~GeV$^2$) data for 
the ${}^{12}$C target in these measurements were used to check 
their  relation to  the BCDMS and CCFR structure functions.   For this, in Ref.~\cite{Fomin:2010ei} the 
extracted  per nucleon $F_{2A}(x,Q^2)$ was corrected for target mass (TM) effects using 
the relation~\cite{Schienbein:2007gr}:
\vspace{-0.2cm}
\begin{align}
  F_{2A}(x,Q^2) &= \frac{x^2}{\xi^2r^3}F_{2A}^{(0)}(\xi,Q^2) + \frac{6M^2x^3}{Q^2r^4} h_2(\xi,Q^2)
  \notag \\ & + \frac{12M^4x^4}{Q^4r^5} g_2(\xi,Q^2),
  \label{fmcorr}
\end{align}
where $h_2(\xi,Q^2) = \int_\xi^A u^{-2} F_{2A}^{(0)}(u,Q^2) du$ and
$g_2(\xi,Q^2) = \int_\xi^A v^{-2}(v-\xi)F_{2A}^{(0)}(v,Q^2) dv$, with the 
Nachtmann variable $\xi = 2x/(1+r)$ and  $r=\sqrt{1+Q^2/\nu^2}$. 
Here,  $F_{2A}^{(0)}(\xi,Q^2)$ is the corrected structure function for which the $Q^2$-dependence 
within the partonic model should come from the evolution equation.  The $h_2$ and  $g_2$ factors have been evaluated assuming a
common $Q^2$ dependence of $F_2^{(0)}$ for all nuclei and simple fit for $F_2^{(0)}(\xi,Q^2_0)$ at $Q_0^{2} = 7$~GeV$^2$.

To relate the extracted $F_{2A}^{(0)}(\xi,Q^2)$ at large $\xi$   
to the  BCDMS and CCFR results, in Ref.~\cite{Fomin:2010ei}  the $Q^2$-dependence of $F_{2A}^{(0)}$
was fit to the world data, including JLab's high-$Q^2\ge 6$~GeV$^2$ data, at several values of $\xi$. 
The functional form of the fit was chosen to have  a $\log Q^2$ term to be consistent with QCD evolution.
Then, using this fit, the  extracted $F_{2A}^{(0)}(\xi,Q_0^2)$ at $Q_0^2=7$~GeV$^2$ was extrapolated
to the BCDMS and CCFR kinematics at large $\xi$.   This extrapolation~\cite{Fomin:2010ei} resulted in 
the slope factor of $s=15\pm 0.5$ for the $^{12}$C target  indicating that 
the JLab data are consistent  with the BCDMS results,  with the latter showing only marginal 
strength of high-momentum component in the nuclear wave function~\cite{Benvenuti:1994bb}~(see above discussion).

However, to have the final answer on the relation  of the JLab structure functions to the  higher-$Q^2$ BCDMS and CCFR data, one 
needs a full account of QCD evolution.  To do so, we derive in the following section the QCD evolution equation for superfast quarks in 
leading order approximation and apply it to $F_{2A}^{(0)}(\xi,Q_0^2)$, to evolve it to BCDMS and CCFR kinematics.

 \medskip


\section{Evolution Equation}
\label{eveq}

We start with the leading order evolution equation for quarks in nuclei:
\begin{eqnarray}
  \frac{d q_{i,A}(x,Q^2)}{ d \log Q^2}
  & = &
  \frac{\alpha_s}{2\pi} \int\limits_{x}^A\frac{dy}{y}
  \left( q_{i,A}(y,Q^2) P_{qq}\left(\frac{x}{ y}\right)
  \right.\nonumber \\
  &  + &
  \left.
  g_A(y,Q^2) P_{qg}\left(\frac{x}{y}\right)\right),
  \label{baseq}
\end{eqnarray}
with the goal of calculating the evolution for the per nucleon structure function $F_{2A}$,
defined at leading order as:
\begin{equation}
  F_{2A}(x,Q^2) = \frac{1}{A}\sum\limits_i e_i^2 x q_{i,A}(x,Q^2),
  \label{F2def}
\end{equation}
where one sums over the flavors of active (anti)quarks.
Note that in Eq.~(\ref{baseq}) the upper limit of the integration is $A$, and thus the integrand in the range of $y>1$ accounts for the 
contribution of the superfast quarks to the evolution of the  partonic distribution $q_{i,A}$ probed at a given $(x,Q^2)$.

Above, the  $q_{i,A}$ functions are  the $i$-flavor quark and antiquark distributions in nuclei, while
$g_A$ represents the nuclear gluon distribution. The splitting functions are:
\begin{eqnarray}
P_{qq}(x) &  = &  C_2\left[(1+x^2)\left(\frac{1}{1-x}\right)_+  + \frac{3}{2} \delta(1-x)\right]\nonumber \\
P_{qg}(x)  & = & T \left[(1-x)^2  + x^2\right],
\label{splitfun}
\end{eqnarray}
with $C_2 = \frac{4}{3}$ and $T = \frac{1}{2}$. Here the $+$ denominator  is the Altarelli-Parisi function, defined as~\cite{Altarelli:1977zs}:
\begin{equation}
\int\limits_{0}^1 dz \frac{f(z)}{(1-z)_+} = \int\limits_0^1 \frac{f(z) -f(0)}{1-z}.
\label{apar}
\end{equation}

We proceed by changing the integration variable in Eq.~(\ref{baseq}) to $z = \frac{x}{y}$ which yields
\begin{eqnarray}
\frac{d q_{i,A}(x,Q^2)}{ d \log Q^2}  & = &  \frac{\alpha_s}{ 2\pi} \int\limits_{{x/A}}^1\frac{dz}{z} \left( q_{i,A}\left(\frac{x}{z},Q^2\right) P_{qq}({z})
  \right.\nonumber \\ 
  &  + & \left.   g_A\left(\frac{x}{z},Q^2\right) P_{qg}(z)\right).
  \label{baseqz}
\end{eqnarray}
Substituting  the splitting functions of Eq.~(\ref{splitfun}) into the above equation results in:
\begin{eqnarray}
 \frac {d q_{i,A}(x,Q^2)}{d \log Q^2} & = &  \frac{\alpha_s}{ 2\pi}\left\{ 2 q_{i,A}(x,Q^2) + \frac{4}{3}\int\limits_0^1dz  
  \frac{f(z)}{(1-z)_+}\right. \nonumber \\
  &  + &  \left. \int\limits_{x/ A}^1dz \frac{(1-z)^2 + z^2}{2 z}g_A\left(\frac{x}{ z},Q^2\right)\right\},
  \label{baseqs}
\end{eqnarray}
where
\begin{equation}
  f(z) = \frac{1+z^2}{z} q_{i,A}\left(\frac{x}{z}, Q^2\right)\theta\left(z-\frac{x}{ A}\right).
\end{equation}
Applying the rule of Eq.~(\ref{apar}) into the second integral of Eq.~(\ref{baseqs}), one obtains the final expression for 
the evolution equation of quarks in the nucleus in the form:
 \begin{eqnarray}
  & & \frac{d q_{i,A}(x,Q^2)}{d \log Q^2}  =    \frac{\alpha_s}{ 2\pi}\left\{ 2\left(1+\frac{4}{3}\log\left(1-\frac{x}{A}\right)\right) q_{i,A}(x,Q^2) \right.
  \nonumber \\
  & & +    \frac{4}{3}\int\limits_{x/A}^1\frac{dz}{1-z}\left(\frac{1+z^2}{ z}q_{i,A}\left(\frac{x}{z},Q^2\right) - 2 q_{i,A}(x,Q^2)\right)\nonumber \\
  & & +   \left. \int\limits_{x/A}^1dz \frac{(1-z)^2 + z^2}{ 2 z}g_A\left(\frac{x}{z},Q^2\right)\right\}.
\end{eqnarray}

This equation can be used to obtain the evolution equation for the structure function $F_{2A}$ defined according to 
Eq.~(\ref{F2def}).   Multiplying both sides above by $e_i^2x$ and summing by contribution of all (anti)quarks
one obtains  the evolution equation for the nuclear structure function $F_{2A}$ in the form
 \begin{eqnarray}
  & & \frac{d F_{2A}(x,Q^2)}{ d \log Q^2}   =    \frac{\alpha_s}{ 2\pi}\left\{ 2\left(1+\frac{4}{3}\log\left(1-\frac{x}{A}\right)\right) F_{2,A}(x,Q^2) \right.
  \nonumber \\
  & &   +    \frac{4}{3}\int\limits_{x/A}^1\frac{dz}{1-z}\left((1+z^2)F_{2A}\left(\frac{x}{z},Q^2\right) - 2 F_{2A}(x,Q^2)\right) \nonumber \\
  & &  +   \left.  \frac{f_Q}{2}\int\limits_{x/A}^1dz [(1-z)^2 + z^2]\frac{x}{z}G_A\left(\frac{x}{z},Q^2\right)\right\},
  \label{F2A_eveq}
\end{eqnarray}
where $f_Q  = \sum\limits_i (e_i^2 + \bar e_i^2)$ and $G_A(x,Q^2) = x g_A(x,Q^2)/A$.  
One interesting property of the above 
equation  which has a nuclear origin is the factor $\log(1-\frac{x}{A})$ which introduces  a  non-trivial A dependence 
into the evolution equation. The effect of this term   can be observed  for light nuclei  at large $x$ kinematics. 


\begin{figure}[ht]
  \centering
  \includegraphics[width=0.5\textwidth]{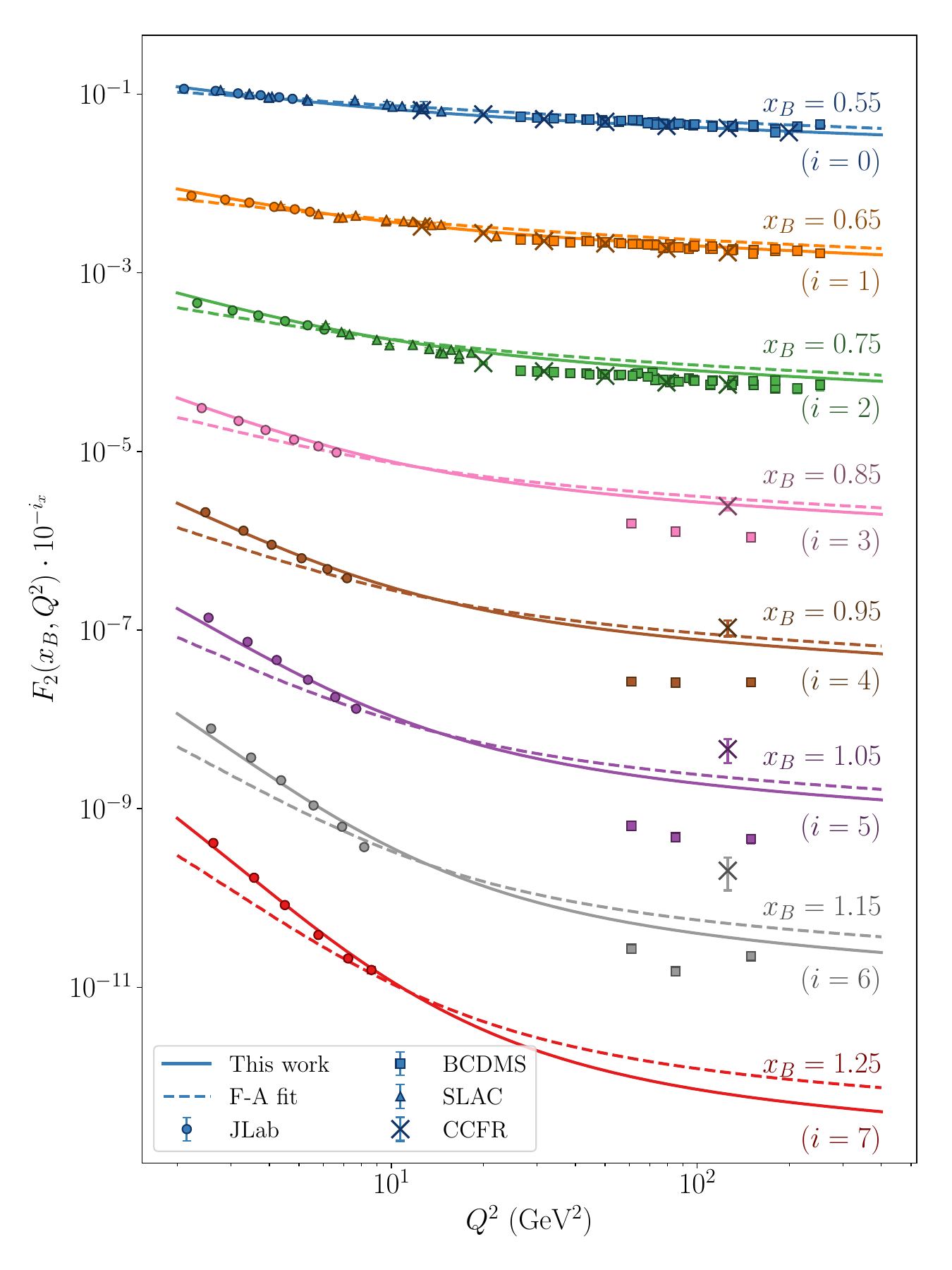}
  \vspace{-0.6cm}
  \caption{
    (Color online.) Comparison of evolution equation results for the per nucleon $F_{2A}$ of $^{12}$C 
    to experimental measurements. 
    The structure function is multiplied by $10^{-i_x}$ in order to separate the curves;
    the values of $i_x$ for each $x$ value are given in the plot.  The solid curves incorporate evolution in the fit (see Sec.~\ref{fit2}), 
    the dashed curves are  the result of QCD evolution in which as an input we used the fit of Ref.~\cite{Fomin:2010ei} (see discussion in Sec.~\ref{eveqnum}).}
\label{fig1}
\end{figure}

\medskip

\section{Evolution of $F_{2A}$ from moderate to high $Q^2$}
\label{eveqnum}
At large $x > 0.1$, we can safely neglect the gluonic distribution $G_A$ in Eq.~(\ref{F2A_eveq}),
after which the evolution of  the structure function $F_{2A}$  at given ($x$,$Q^2$) will be defined by the same structure 
function at $x^\prime \ge x$ and some initial $Q^2_0$.
Such a situation  allows us to relate the  $F_{2A}$ structure  functions at  high $Q^2$ (BCDMS and  CCFR) kinematics 
to the same structure function at moderate-$Q^2$ (JLab) kinematics using Eq.~(\ref{F2A_eveq}),
without requiring the knowledge of  the nuclear gluonic distribution $G_{A}$.
 
To do so, first, we  use as an input to Eq.~(\ref{F2A_eveq}) the 
same parametrization of $F^{(0)}_{2A}(\xi,Q_0^2)$ at $Q_0^2=7$~GeV$^2$~\cite{AFpc}
for the $^{12}$C nucleus that was used in the high-$\xi$ and  high-$Q^2$  extrapolation of 
Ref.~\cite{Fomin:2010ei} (referred to hereafter as QCD evolution with F-A fit).
With this input, Eq.~(\ref{F2A_eveq}) is solved numerically, covering the $Q^2$ range of $2$-$300$~GeV$^2$.
The TM-uncorrected  $F_{2A}$ is then obtained from $F_{2A}^{(0)}$ by reintroducing target mass effects according to Eq.~(\ref{fmcorr}).

\begin{figure}[t]
\vspace{-0.2cm}
  \centering
  \includegraphics[width=0.5\textwidth]{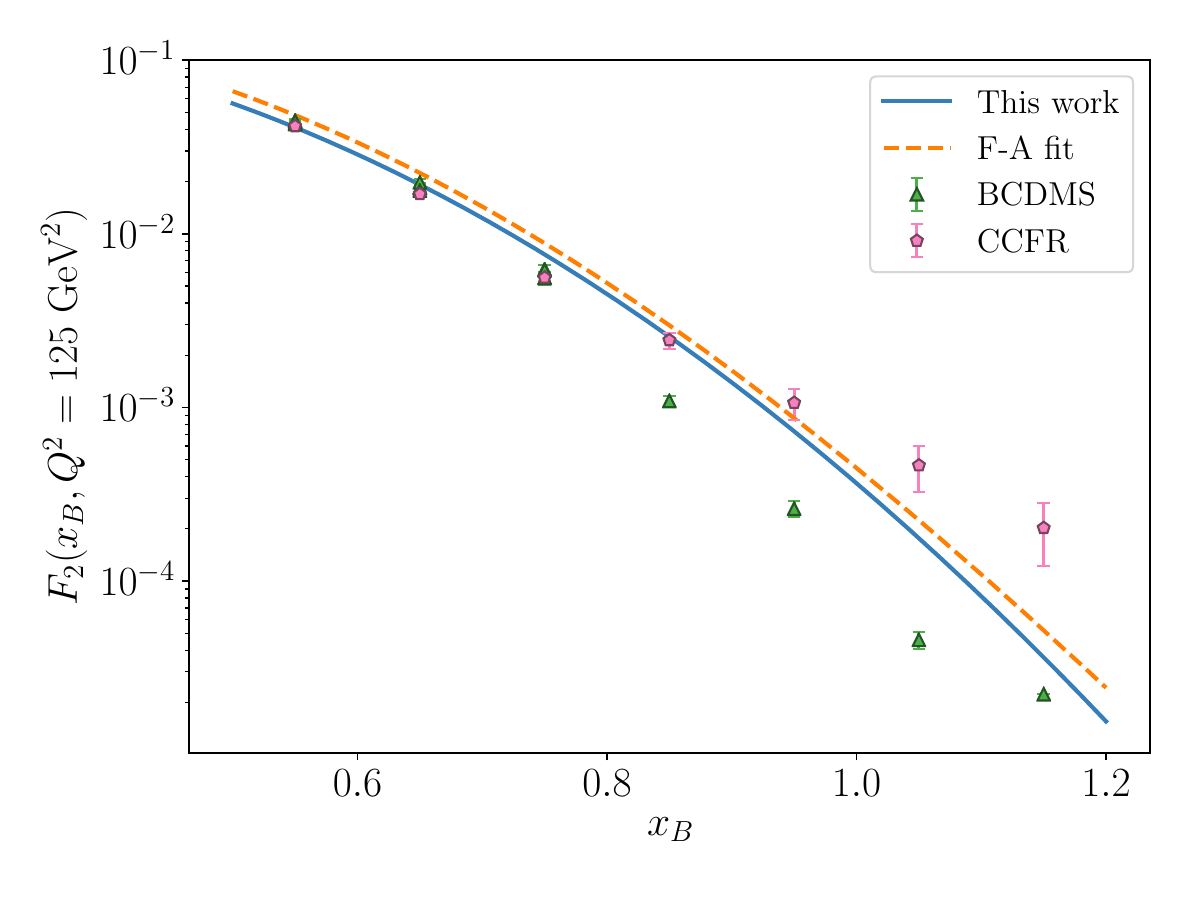}
\vspace{-0.5cm}
  \caption{The $x$ dependence of $F_{2A}$ at $Q^2 = 125$~GeV$^2$.  The solid and dashed curves are the same as in Fig.\ref{fig1}.}
  \label{xdep}
\end{figure}

The result of the calculations is given by the dashed curves  in Figs.~\ref{fig1} and \ref{xdep}, along with experimental data and SLAC ``pseudodata.''
The JLab~\cite{Fomin:2010ei} and BCDMS~\cite{Benvenuti:1987zj,Benvenuti:1994bb}  data are  measurements of the structure function per nucleon,
whereas the SLAC pseudodata are obtained  according to Ref.~\cite{Fomin:2010ei}
by multiplying deuteron $F_2$ measurements~\cite{Whitlow:1991uw} by the EMC ratio  measured in Ref.~\cite{Gomez:1993ri}.
The CCFR data at $x>0.75$  were given without an absolute normalization~\cite{Vakili:1999qt},
so in Fig.~\ref{fig1} the $x = 0.75$ point was normalized to the previous CCFR measurement at $x\le 0.75$, for which the absolute 
values have been measured~\cite{Seligman:1997mc}.
Note that the discrepancy between the dashed curves in Fig.~\ref{fig1} and the low-$Q^2$ JLab data is due to the fact that F-A parameterization is 
fitted in the $6\le Q^2 \le 9$~GeV$^2$ region only.

As the figure shows, the  F-A  parameterization extended to the high-$Q^2$ domain of the CCFR and BCDMS experiments ($Q^2 \sim 125$~GeV$^2$) through 
QCD  evolution does not prefer the BCDMS data as the phenomenological $Q^2$ extrapolation of Ref.~\cite{Fomin:2010ei} had indicated. 
In fact, QCD evolution of JLab data shows better agreement with the CCFR data at $x\le 1.05$,  and results  in a slope factor 
$s=13\pm0.4$ for the range of $0.75 \le x < 1.25$.

\medskip

\section{The $\xi$ Parameter Fitting of JLab Data}
\label{fit2}

Even though QCD evolution of the F-A parameterization predicts a softer
$x$ dependence for $F_{2A}(x)$ at $Q^2=125$~GeV$^2$ than the extrapolation
quoted in Ref.~\cite{Fomin:2010ei} ($s=13\pm0.4$, compared to $s=15\pm 0.5$),
it overestimates the $F_{2A}$  data at $x\le 0.75$  and $Q^2\ge 20$~GeV$^2$ where 
structure functions are reliably measured (see the three data sets and the dashed curves in the upper part of Fig.~\ref{fig1}).  
Additionally the QCD evolution  underestimates the  $F_{2A}$ data at higher $x\ge 0.85$
and $Q^2\le 5$~GeV$^2$ (see dashed curves in Fig.~\ref{fig1}).
In the latter case,  the underestimation at low $Q^2$ is due to the fact that only $Q^2\ge 6$~GeV$^2$
data have been used to fit the extracted structure function $F_{2}^{(0)}$ in the F-A parameterization.
The other discrepancies can be attributed to the specific model of target mass
corrections adopted in Ref.~\cite{Fomin:2010ei}
({\sl cf.}\ Eq.~(2),  as well as Ref.~\cite{Schienbein:2007gr}).
As was discussed in the previous section, after applying QCD evolution to  the  $F^{(0)}_{2A}$  structure function
the target mass effects are reapplied to compare the evolved results with the empirical data.
We find that the $Q^2$ dependence introduced by the factor of $x^2/(\xi^2r^3)$
in Eq.~(\ref{fmcorr}) partially cancels out the $Q^2$
dependence introduced by evolution, thus giving  the  final result a softer
$Q^2$ dependence.
 
To address the problem of these discrepancies we consider a different approach to target mass corrections. 
In the new approach
the Nachtmann variable $\xi$ is  treated as a  scaling parameter, representing
the light cone momentum fraction variable instead of $x_B$. Within such an approach,
$\xi$ enters into the QCD evolution equations, and no  additional target mass corrections are applied
to the data.  It is worth mentioning that such an approach is justified at leading order, where
$\xi$-scaling corresponds to the target mass correction in the collinear approximation~\cite{Aivazis:1993kh}.
That such an approach is justified follows also from the empirical observation in
Ref.~\cite{Fomin:2010ei}  that  the raw (uncorrected) $F_{2A}$ data plotted as a
function of $\xi$ exhibit better scaling properties than the data  corrected
according to Eq.~(\ref{fmcorr}).

Within such an approach  we analyzed the uncorrected JLab data
considering the structure function as a function of $\xi$ and attempting to parameterize 
it  in the form~\cite{Accardi:2009br}:
\begin{equation}
  F_{2A}(\xi,Q^2)
  =
  F_{2A}^{\mathrm{LT}}(\xi,Q^2)
  \left(1 + \frac{c_1 \xi^{c_2}(1+c_3\xi)}{Q^2}\right),
  \label{fitform}
\end{equation}
where the "LT" indicates the leading twist contribution to the structure function, which can be used as an input 
for the evolution equation. The latter is parametrized at an initial scale $Q_0^2=\sqrt{18}~\text{GeV}^2$ as
\begin{equation}
  F_{2A}^{\mathrm{LT}}(\xi,Q_0^2) = \exp(p_0 + p_1\xi + p_2\xi^2)
  \label{LTform}
\end{equation}
in the range  of $0.5 < \xi < 1.3$.
The value of $F_{2A}$ at other scales is obtained by applying the evolution
equation of Eq.~(\ref{F2A_eveq}) to Eq.~(\ref{LTform}).
To fit the parameters of Eqs.~(\ref{fitform}) and (\ref{LTform}),
we used all the JLab data  with $x>0.5$.

We employed three different strategies to perform the fit.
The first was to use differential evolution~\cite{Storn1997}, a
multidimensional optimization method in which a population of candidate
solutions can mutate and evolve, and in which the population members with
the best ``fitness'' ({\sl e.g.}, the lowest $\chi^2$ values) are combined
to produce new candidate solutions.
In this, we use the $\chi^2$ of the fit as the fitness function.
The second strategy was to use the standard MINUIT2 library functions with
a $\chi^2$ fit function.
Lastly, the third was a bootstrap method, in which we generated populations
by sampling the data points from a Gaussian with a center and width determined
by their experimental values and statistical errors.
For each of these populations, a $\chi^2$ fit was performed using MINUIT2,
and subsequently the distributions of the fit parameters were used to
determine their averages and standard deviations.
In all three cases, the fitness parameter ($\chi^2$) was determined using
only the statistical, and not the systematic, errors of the data reported in
Ref.~\cite{Fomin:2010ei}, as the systematic errors are dominated by beam energy
and detector setting uncertainties, and are hence expected to be highly correlated.

Using all three strategies, we first performed fits to the full six-parameter
form of Eq.~(\ref{fitform}). We then performed fits without a higher-twist
correction, {\sl i.e.}, with the form of Eq.~(\ref{LTform}) only.
We found with all three strategies that the six-parameter fit did not
yield significant improvement in the $\chi^2$ value compared to the
three-parameter fit.
Moreover, in the six-parameter fit, the central parameter values varied wildly
with small changes in the data set used for the fit, but generally preferred
small values of $c_1$.
On the other hand, the three-parameter fit without the HT factor yielded very
robust results for the parameters, with central values, standard errors,
and covariances comparable between the three approaches.
We therefore select the three parameter fit as the optimal one.

\begin{table}
  \centering
  \caption{
    Parameters found in the three-parameter fit
    by the three fitting strategies,
    along with their standard errors.
  }
  \bgroup
  \def\arraystretch{1.5}
  \begin{tabular}{
      c @{\hspace{1em}} c @{\hspace{1em}} c @{\hspace{1em}} c @{\hspace{1em}}
    }
    \hline
    ~ & $p_0$ & $p_1$ & $p_2$
    \\ \hline \hline
    Diff.\ Evol.
    & $ 0.248 \pm 0.005$
    & $ 4.42 \pm 0.01$
    & $-9.15 \pm 0.01$
    \\
    MINUIT2
    & $ 0.235 \pm 0.006$
    & $ 4.45 \pm 0.02$
    & $-9.17 \pm 0.01$
    \\
    Bootstrap
    & $ 0.235 \pm 0.005$
    & $ 4.45 \pm 0.01$
    & $-9.17 \pm 0.01$
    \\ \hline
  \end{tabular}
  \egroup
  \label{table:pars}
\end{table}

The results for the parameters for $Q^2_0 = \sqrt{18}$~GeV$^2$, along with their standard errors are 
presented in Tab.~\ref{table:pars}.
 
\begin{figure}[ht]
  \centering
  \includegraphics[width=0.5\textwidth]{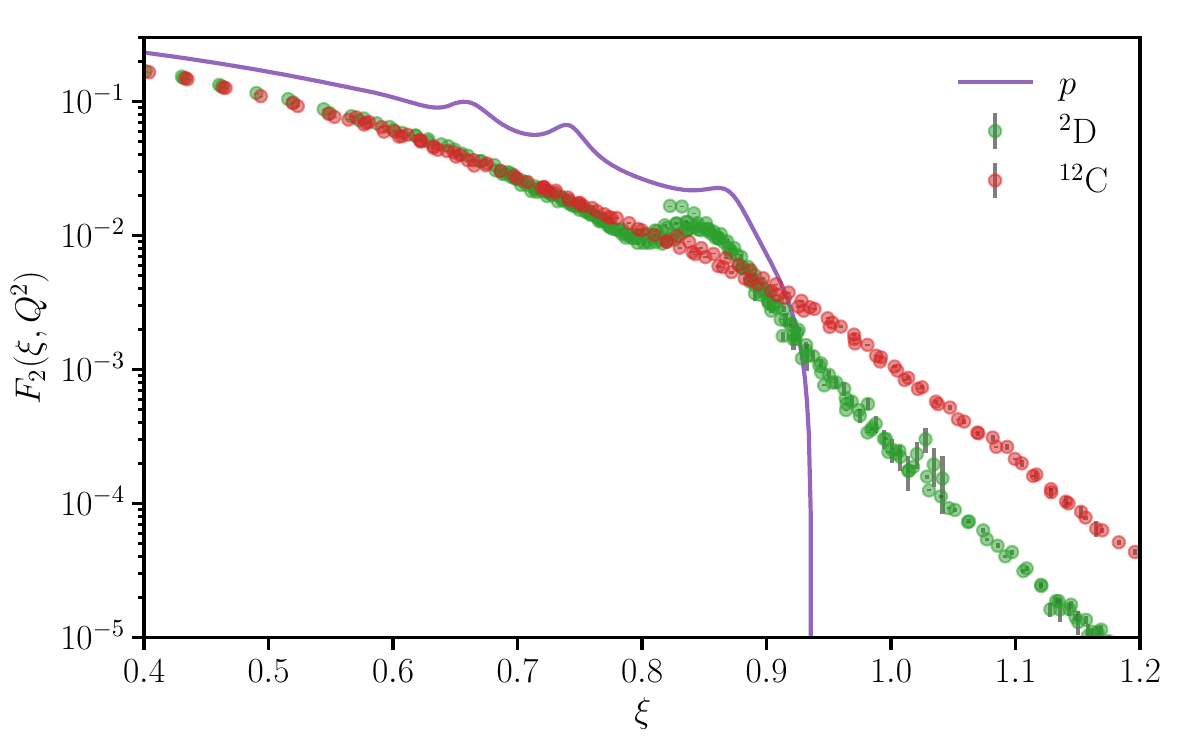}
  \caption{
    (Color online.)
    World data of $F_{2A}$ for $^{12}$C and the deuteron as a
    function of Nachtmann variable $\xi$ for
    $3.5~\text{GeV}^2 < Q^2 < 4.5~\text{GeV}^2$~\cite{
      Schutz:1976he,Rock:1991jy,Lung-thesis:1992,
      Arrington:1998ps,Fomin:2010ei}.
    Solid curve shows $F_{2p}(\xi,Q^2=4~\text{GeV}^2)$
    using the parameterization of Ref.~\cite{Bodek:1979rx}.
  }
\label{fig:smearing}
\end{figure}

Our observation of the negligible contribution from the higher twist effects can be understood based on 
a combination of quark-hadron duality
and Fermi motion effects  which results in a nearly-complete cancellation of the
higher-twist effects for the $^{12}$C nucleus (see also Ref.\cite{Moffat:2019qll}).
Usually, quark-hadron duality for the proton structure function is observed when
the structure function is smeared over some range of final produced mass
$W_N$ (see {\sl e.g.}~\cite{Melnitchouk:2005zr}).
For a nuclear target, this smearing is inherently accomplished by the Fermi
motion of the nucleons within the nucleus.
To demonstrate this, we compare in Fig.~\ref{fig:smearing} the $F_2$ structure
functions for  $Q^2=4$~GeV$^2$ (which is close to our choice of $Q^2_0$)  for the deuteron, $^{12}$C,
and a phenomenological parametrization of the proton~\cite{Bodek:1979rx}.
While one  observes resonance structures in the proton  and deuteron $F_2$ structure functions,
these effects are significantly suppressed in the $^{12}$C data.
We expect that this effect will be even more significant for heavier nuclei,
which gives a new possibility for quality fitting of nuclear DIS structure
functions at high $x$.

With the parameters quoted in Tab.~\ref{table:pars}, we have reconstructed the
leading-twist structure function $F_{2A}^{\mathrm{LT}}(\xi,Q^2)$ according to
Eq.~(\ref{LTform}) at $Q^2_0 = \sqrt{18}$~GeV$^2$, and evolved it to all other
$Q^2$ using the evolution equation (\ref{F2A_eveq}).
With this procedure, we calculate the $F_{2A}$ structure function at CCFR and BCDMS
kinematics.  In Figs.~\ref{fig1} and \ref{xdep}, the solid curves represent the
results of this calculation. The parameter errors were also propagated into
$F_{2A}$ at these kinematics and included as shaded bands in the plot, but these
bands cannot be seen because they are smaller than the line width of the curves.
(Note the small standard errors quoted in Tab.~\ref{table:pars}).
As the comparison shows, QCD evolution now describes the $x=0.55$ and $0.65$
data at high $Q^2$ very well, while slightly overestimating the $x=0.75$ data
at high $Q^2$. Note that the dashed curves in Figs.~\ref{fig1} and \ref{xdep} represent 
the result of  QCD evolution in which as an input we used the F-A fit at $Q^2_0 = 7$~GeV$^2$ from Ref.\cite{Fomin:2010ei}.

For the slope factor, we obtain $s = 13.0\pm 1.1$ for $x\ge 0.75$ and $Q^2=125$~GeV$^2$. This result 
is practically the same one obtained  from 
evolution of the F-A parametrization. Thus one concludes that our overall result for the nuclear structure function $F_2(x,Q^2)$ 
is somewhat between the
CCFR ($s\approx 8.3$) and BCDMS ($s\approx 16.5$) estimates,
while the absolute magnitude of $F_{2A}$ is closer to the CCFR data at
$x\le1.05$. Remind that  phenomenological $Q^2$ extrapolation of $F-A$ parameterization\cite{Fomin:2010ei}  resulted in 
the slope factor $s=15\pm 0.5$  favoring the BCDMS result.
 
\section{QCD Evolution Based  Fit of $F_{2A}(\xi,Q^2)$}
\label{fit_of_fits}
The success of the QCD evolution equation in describing the structure function data below and above $Q_0^2 = \sqrt{18}$~GeV$^2$ motivates us in 
presenting $F_{2A}$ in a parametric form  that covers the whole considered $Q^2$ range starting $Q^2\ge 2$~GeV$^2$ and $x>0.5$.  Such a fit 
can be used for evaluating nuclear DIS cross sections in a wide range of kinematics relevant for 12 GeV JLab and  EIC physics.

\begin{figure}[ht]
  \centering
  \includegraphics[width=0.5\textwidth]{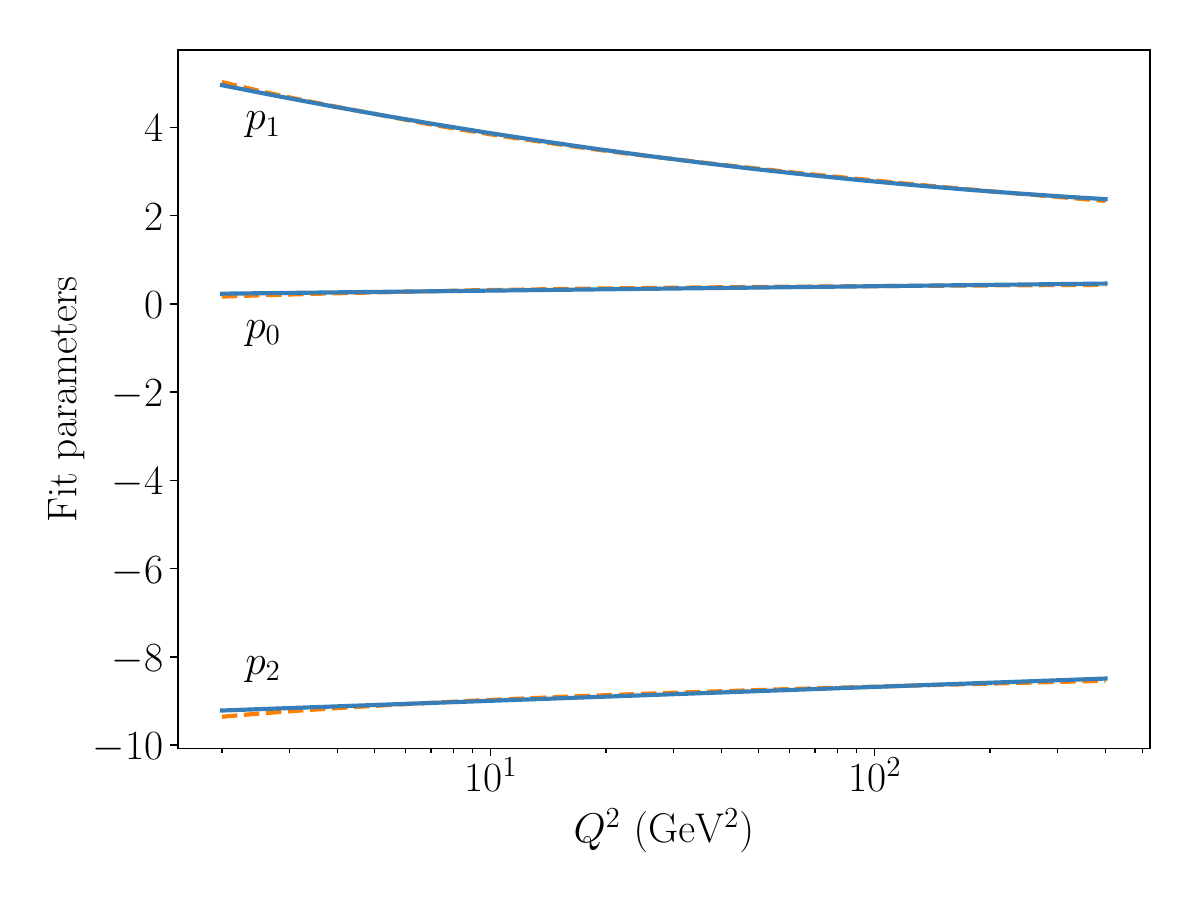}
  \caption{
    (Color online.)
    The fit of $p_0$, $p_1$ and $p_2$ as a function $t = \log{Q^2\over 1\mathrm{GeV}^2}$. Dashed  curves are the results generated by 
    evolution equation and solid curves correspond to the polynomial fit of Eq.~(\ref{par_pars}).
  }
\label{fig:par_pars}
\end{figure}

In performing such a fit we again used the analytic form of Eq.~(\ref{LTform}),  where the parameters $p_0$, $p_1$ and $p_2$ are determined on a per $Q^2$ value basis
by fitting the values of $F_{2A}$ as determined by QCD evolution.  Because of the QCD evolution,
these parameters are inherently $Q^2$ dependent,
and we express this dependence  in a simple polynomial fit in the variable
$t = \log{Q^2\over 1\mathrm{GeV}^2}$  as follows:
\begin{eqnarray}
p_0(t) & = &  a_0 + b_0 t \nonumber \\
p_1(t) & = & a_1 + b_1 t + c_1 t^2 \nonumber \\
p_2(t) & = & a_2 + b_2 t.
\label{par_pars}
\end{eqnarray}
The central values of the $a_i$ and $b_i$ parameters are presented in Table \ref{table:par_pars}. Fig.~\ref{fig:par_pars} also shows both the
$t$ dependence of the $p_0$, $p_1$ and $p_2$ parameters and the results of the polynomial fit.
\begin{table}[ht]
  \centering
  \caption{Parameters defining the $t$ dependence of $p_0(t)$, $p_1(t)$ and $p_2(t)$ function in Eq.\ref{par_pars}.}
\vspace{0.2cm}
  \begin{tabular}
{    c @{\hspace{1em}} c @{\hspace{1em}} c @{\hspace{1em}} c @{\hspace{1em}}  c @{\hspace{1em}} c @{\hspace{1em}} c @{\hspace{1em}} }
    \hline
     $a_0$ & $b_0$ & $a_1$ & $b_1$ & $c_1$ & $a_2$ & $b_2$ \\ 
    \hline
     0.201 &  0.043 & 5.504 & - 0.828 & 0.051 & -9.309  & 0.137 \\
      \hline
  \end{tabular}
  \label{table:par_pars}
\end{table}
Here one observes very smooth $t$ dependence  consistent with 
the above observation of negligible   higher twist effect for nuclear $F_{2A}$.
We expect this parametrization of $F_{2A}$ to be valid for $Q^2$ up to 400~GeV$^2$,
the maximum value to which we performed QCD evolution,
and it gives a simple way of estimating cross sections for deep inelastic scattering
in the superfast quark region.

\medskip


\begin{figure}[ht]
  \centering
  \includegraphics[width=.5\textwidth]{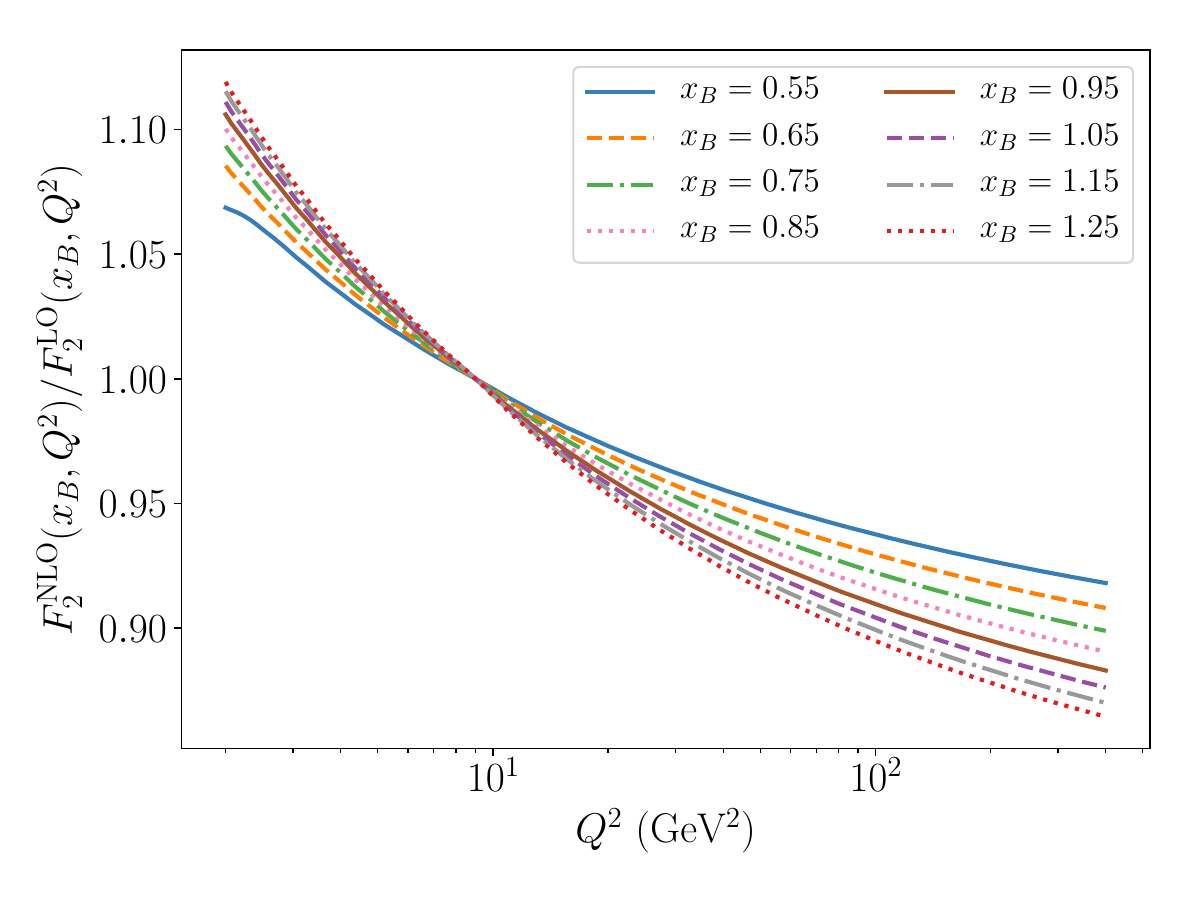}
  \vspace{-0.6cm}
  \caption{
    (Color online.)
    Ratio of $F_{2A}$ calculated for $^{12}$C
    using NLO evolution to LO evolution.
  }
  \label{fig:NLOLO}
\end{figure}

\section{NLO Corrections}
\label{NLO}

To estimate the accuracy of the leading order (LO)
evolution equation presented in Figs.~\ref{fig1} and \ref{xdep},
we have also evolved the fit of $F_{2A}$ obtained in Sec.~\ref{fit2}
at next-to-leading order (NLO).

To perform NLO evolution, we make similar approximations to those described
in Sec.~\ref{eveq}. We neglect the gluon distribution, since this is small
at $x > 0.2$. At NLO, the quark splitting functions have non-diagonal terms,
and the splitting functions for singlet and non-singlet mixtures become
different (see, {\sl e.g.}, \cite{Ellis:1991qj}).
We take advantage of the fact that $^{12}$C is isospin symmetric,
and evolve $F_{2A}$ as a singlet distribution
(within the approximation where gluons are neglected).

Additionally, at NLO, Eq.~(\ref{F2def}) is no longer exact,
but $F_{2A}$ must be determined from the quark distributions
through a Mellin convolution with the NLO Wilson coefficients.
One can still evolve $F_{2A}$ directly, however, by folding these
Wilson coefficients into the splitting functions.
In Ref.~\cite{Miyama:1995bd}, this is described as a ``one-step'' method.
We perform such a one-step method in our NLO evolution of $F_{2A}$.

Since we are using NLO evolution primarily to estimate the accuracy of LO
evolution, we present in Fig.~\ref{fig:NLOLO} the ratio of NLO-evolved
to LO-evolved $F_{2A}$, with the parametrization (\ref{LTform})
and the parameters in Tab.~\ref{table:par_pars} at $Q_c^2=9$~GeV$^2$
as a common starting point.  The choice of $Q_c^2$ is justified by the fact that it 
corresponds to the largest $Q^2$  data measured at JLab experiment and 
we achieved a reasonable description of the $F_2(x,Q_c^2)$ extracted from 
these data.
All lines thus intersect in the figure  at $Q^2 = Q_c^2$, with a ratio of $1$.
One can see from this figure that the amount of evolution that occurs
is enhanced by NLO corrections,
and this enhancement results in a  greater suppression of $F_{2A}(x,Q^2)$ for larger $x_B$.
In fact, when $Q^2\sim 125$~GeV$^2$, NLO corrections are as much as $11\%$.
Such a correction however does not alter our conclusion that the QCD evolution of JLAB data 
results in a $F_{2A}$ that favors CCFR at $x\le 1.05$ and  predicts magnitudes somewhat in the middle of 
CCFR and BCDMS data at $x\ge 1.15$.

However, NLO corrections can be  sizable enough that they will be
necessary to account for to make precision predictions in
larger-$Q^2$ regions relevant to  the LHC and the anticipated EIC kinematics.
A detailed study of NLO evolution to such high-$Q^2$ regimes
will be performed in a future work.

\medskip


\section{Summary and  Conclusions}
\label{SumCon}
We derived the evolution equation for superfast quarks in nuclei in the leading order approximation.
For  the $F_{2A}$ structure function at high $x$,
in an approximation in which  the gluon distribution is  neglected,
QCD evolution allows high-$Q^2$ values of $F_{2A}$ to be determined
by the same $F_{2A}$ measured at some initial value of $Q_0^2$.
Using this property and the  parameterization of  $F_{2A}$  at moderate $Q^2=\sqrt{18}$~GeV$^2$,
we fit a parametric form to the Jefferson Lab data and 
used the evolution equation to calculate $F_{2A}$ in the range of $60 < Q^2 < 200$~GeV$^2$,
at which the previous measurements of  superfast quark distributions  have been made.
Our approach uses the  QCD evolution equation directly to determine  nuclear structure functions $F_{2A}$ at large $x$. 
This approach has an advantage over modeling  of nuclear structure functions based on a convolution of the free
nucleon $F_{2N}$ structure function and nuclear dynamics. In the latter case one deals with uncertainties  inherent to the models, where different nuclear 
effects such as Fermi motion of nucleons, medium modification of nucleon PDFs and possible final state interactions should be taken into 
account.

Our calculation demonstrates that the JLab high-precision, moderate-$Q^2$ measurement of the 
${}^{12}$C  structure function is 
in better agreement with the CCFR data at $Q^2=125$~GeV$^2$ and $x\le 1.05$ with the slope factor $s$ indicating 
a sizable  contribution of the high-momentum nuclear component in the generation of superfast quarks.  Our results  at $x>1.05$  is somewhat in 
the middle of CCFR and BSDMS results of nuclear structure function data.

\medskip

\begin{acknowledgments}
We are thankful to Drs. John Arrington and  Nadia Fomin for numerous  discussions and providing results of their analysis and the  fit 
of the Carbon structure  function of their Jefferson Lab experiment, and to Drs. Matthew Dietrich and Jannes Nys for discussions on fitting strategies.    This  work is supported by U.S. DOE grant under contract DE-FG02-01ER41172.  Also, AF was supported by the U.S. Department of Energy, Office of Science, 
Office of Nuclear Physics, contract no. DE-AC02-06CH1135 and an LDRD 
initiative at Argonne National Laboratory under Project No. 2017-058-N0.

\end{acknowledgments}


 \end{document}